# Magnetic History and Field Sweep Rate Effects on Transformation Hysteresis in $Ni_{41}Co_{9}Mn_{31.5}Ga_{18.5}$

T. Kihara[1]*, T. Miyakawa[2, 3], D. Imatomi[2], X. Xu[2], Y. Ishii[4], H. Hayashi[4, 5], Y. H. Matsuda[4], Y. Adachi[6], and T. Kanomata[7]

[1] *Research Institute for Interdisciplinary Science, Okayama University, Okayama, Okayama 700-8530, Japan*

[2] *Department of Materials Science, Tohoku University, Sendai, Miyagi 980-8579, Japan*

[3] *J-PARC Center, Japan Atomic Energy Agency, Tokai, Ibaraki 319-1195, Japan*

[4] *Institute for Solid State Physics, The University of Tokyo, Kashiwa, Chiba 277-8581, Japan*

[5] *Laboratoire National des Champs Magnétiques Intenses (LNCMI-EMFL), CNRS-UGA-UPS-INSA, Toulouse 31400, France*

[6] *Graduate School of Science and Engineering, Yamagata University, Yonezawa, Yamagata 992-8510, Japan*

[7] *Research Institute for Engineering and Technology, Tohoku Gakuin University, Sendai, Miyagi 984-8588, Japan*

**Abstract:** Hysteresis in magnetic-field-induced structural phase transformations is a major obstacle to the application of metamagnetic shape memory alloys. One promising approach to reduce hysteresis is to utilize a metastable martensitic state generated by magnetic-field cycling. However, the mechanism by which the metastable state reduces hysteresis remains unclear. Here, microsecond pulsed-field magnetization measurements are used to investigate the transformation kinetics of $Ni_{41}Co_{9}Mn_{31.5}Ga_{18.5}$. Magnetic-field cycling reduces the hysteresis width by approximately 50–60%, demonstrating that substantial suppression persists on a microsecond timescale. Comparison with millisecond pulsed-field measurements suggests that the reverse martensitic transformation field is more sensitive to the field sweep rate for the metastable martensite than for the initial martensite. Possible explanations include enhanced thermal activation associated with a lower effective activation barrier and differences in the athermal contribution arising from different metastable martensitic microstructures.

*Address all correspondence to: tkihara@s.okayama-u.ac.jp

## Introduction

Magnetic-field-induced structural phase transitions provide a rich platform for multifunctional materials because an external magnetic field can simultaneously control magnetic, structural, and electronic properties [1–9]. In metamagnetic shape memory alloys, the coupling between magnetism and the martensitic transformation enables magnetic-field-induced shape recovery [10] and large inverse magnetocaloric effects [4, 11–13], making these materials attractive for magnetic actuators and solid-state refrigeration. From a fundamental perspective, magnetic-field-induced martensitic transformations (MFIMTs) are an attractive subject for studying the interplay among magnetic, structural, and electronic degrees of freedom. However, hysteresis accompanying first-order transformations causes irreversible energy dissipation and increases the magnetic field range required for cyclic operation, limiting practical applications. Because this hysteresis is rooted in the metastability and energy barriers associated with the first-order transformation, its suppression is a fundamental challenge. Reducing this hysteresis therefore requires control of the transformation pathways.

Several approaches address this problem by controlling the crystal structures and microstructures that govern transformation barriers [14–16]. Improving crystallographic compatibility between austenite and martensite through compositional tuning can reduce the elastic cost of forming the phase boundary and substantially narrow the hysteresis [17]. Microstructural control provides another route to reducing hysteresis. In Ni–Mn–In-based alloys, small thermal hysteresis is associated with a relatively uniform microstructure and austenite nucleation at twin boundaries. In contrast, an Fe-doped alloy exhibited nanoscale Fe-rich precipitates and spatial variations in chemical order, which were associated with a nonuniform transformation and larger thermal hysteresis [18]. More broadly, $Eu_2In$ exhibits a first-order magnetoelastic transition with large latent heat and negligible hysteresis, accompanied by very small lattice discontinuities without a change in crystal symmetry [14]. Although this transition differs from a martensitic transformation, it demonstrates that a large latent heat can coexist with negligible hysteresis [14].

Another approach is to use metastable states formed by thermal and magnetic field cycling [19, 20]. $Ni_{50-x}Co_xMn_{31.5}Ga_{18.5}$ alloys ($x = 9$ and 9.7) undergo a martensitic transformation near room temperature on cooling from cubic austenite (A) to the non-modulated 2M martensitic ground state (M) [19]. Applying a magnetic field to M at low temperatures induces the reverse transformation from M to A. In these materials, magnetic-field cycling dramatically decreases the transformation hysteresis at low temperatures. This reduction has been attributed to the formation of the metastable martensitic state ($M^*$). However, the mechanism underlying this hysteresis reduction, by which the $M^*$ modifies the transformation kinetics and the associated energy barriers, remains unclear.

Measurements of MFIMTs at different magnetic-field sweep rates are important for probing these transformation kinetics. Faster field sweeps reduce the time available for thermally activated nucleation and motion of the austenite-martensite interface. This can shift the transformation to higher fields during field increase and to lower fields during field decrease, thereby increasing the hysteresis. In a phenomenological framework based on Seeger's theory, magnetic-field hysteresis is described in terms of athermal and thermally activated contributions. The thermally activated contribution depends on both temperature and magnetic field sweep rate [21, 22]. Experimentally observed sweep-rate dependences of the transformation fields and hysteresis are consistent with this framework [23–25]. Furthermore, analysis of the temperature and sweep-rate dependence of the transformation field using this model enables us to estimate effective activation energies [23]. Therefore, comparing the sweep-rate dependence of the M-to-A and $M^*$-to-A transformations provides insight into the influence of the metastable martensite on transformation kinetics and its role in the hysteresis suppression in Ni–Co–Mn–Ga alloys.

In the present study, magnetization measurements of $Ni_{41}Co_9Mn_{31.5}Ga_{18.5}$ using microsecond single-turn-coil (STC) pulses, with sweep rates of approximately $10^7$ T/s, are compared with previous millisecond nondestructive (ND) pulsed-field measurements [19]. The results demonstrate that substantial hysteresis suppression persists under microsecond STC pulses. Analysis within the phenomenological framework described above examines the thermally activated and athermal contributions to the transformation field, considering possible differences in the metastable states formed under STC and ND fields and their evolution with temperature.

## Experimental methods

Polycrystalline ingots of $Ni_{41}Co_9Mn_{31.5}Ga_{18.5}$ were prepared by arc melting, annealed at 800 °C for 72 h, and quenched in cold water. The ingots were crushed and sieved to obtain powder particles with sizes of 45–63 μm. To relieve the lattice strain induced by crushing, the powder was annealed again at 800 °C for 2 min and then quenched in cold water. Magnetization measurements were performed in destructive pulsed high magnetic fields generated by an STC system at the Institute for Solid State Physics, the University of Tokyo. The powder was mixed with Stycast 1266 epoxy at a powder-to-epoxy mass ratio of 1:1 to suppress eddy-current heating and reduce temperature changes associated with the magnetocaloric effect during the STC pulses.

**Results**

Figure 1(a) compares representative magnetic-field profiles generated by the STC and ND systems. The STC pulse reaches approximately 80 T at 5.3 μs and subsequently reverses polarity, reaching approximately −36 T at 12.2 μs, as indicated by the bottom time axis. The ND pulse reaches approximately 60 T at 17.0 ms, as indicated by the top time axis. The corresponding sweep rates are shown in Fig. 1(b). The sweep rate, defined as $r = |d(\mu_0 H)/dt|$, is of the order of $10^7$ T/s for the STC measurements and $10^3$–$10^4$ T/s for the ND measurements near the transformations.

Figure 2 compares the magnetization curves and their field derivatives obtained using the STC system at 13 K and the ND system at 4.2 K [19]. The solid blue curves represent the first field application after zero-field cooling, whereas the dashed red curves represent the subsequent second field application at the same respective initial temperatures. The magnetization jumps associated with the MFIMTs are observed in both the STC and ND measurements [Figs. 2(a) and 2(c)]. The characteristic transformation fields were determined from the peak positions in $dM/d(\mu_0 H)$ [Figs. 2(b) and 2(d)]. Here, $\mu_0 H_\uparrow$ and $\mu_0 H_\downarrow$ denote the martensite-to-austenite transformation field on increasing field and the austenite-to-martensite transformation field on decreasing field, respectively. These fields are marked by open inverted triangles and open circles. During the first STC field application at 13 K, $\mu_0 H_{\uparrow,\mathrm{STC}}^{\mathrm{1st}} =$ 38.5 T and $\mu_0 H_{\downarrow,\mathrm{STC}}^{\mathrm{1st}} =$ 13.8 T. During the second application, $\mu_0 H_\uparrow$ decreases substantially to $\mu_0 H_{\uparrow,\mathrm{STC}}^{\mathrm{2nd}} =$ 26.1 T, whereas $\mu_0 H_\downarrow$ increases only slightly to $\mu_0 H_{\downarrow,\mathrm{STC}}^{\mathrm{2nd}} =$ 14.3 T. Defining the hysteresis width as $W = \mu_0 (H_\uparrow - H_\downarrow)$, we obtained $W_{\mathrm{STC}}^{\mathrm{1st}} =$ 24.7 T and $W_{\mathrm{STC}}^{\mathrm{2nd}} =$ 11.8 T. Thus, the hysteresis width at 13 K decreases by approximately 52% from the first to the second field application.

The ND measurements exhibit the same qualitative history dependence [Figs. 2(c) and 2(d)]. $\mu_0 H_\uparrow$ decreases from $\mu_0 H_{\uparrow,\mathrm{ND}}^{\mathrm{1st}} =$ 39.0 T to $\mu_0 H_{\uparrow,\mathrm{ND}}^{\mathrm{2nd}} =$ 20.4 T, whereas $\mu_0 H_\downarrow$ decreases only slightly from $\mu_0 H_{\downarrow,\mathrm{ND}}^{\mathrm{1st}} =$ 16.8 T to $\mu_0 H_{\downarrow,\mathrm{ND}}^{\mathrm{2nd}} =$ 16.4 T. The latter change is smaller than the transformation width. In both the STC and ND measurements, the hysteresis reduction therefore mainly results from a decrease in $\mu_0 H_\uparrow$.

Figure 3 summarizes the characteristic transformation fields. The bars in Fig. 3 indicate the widths of the corresponding peaks in $dM/d(\mu_0 H)$, representing the field ranges over which the transformations occur. The $\mu_0 H_\uparrow$ and $\mu_0 H_\downarrow$ measured using STC show similar overall temperature dependence. Both $\mu_0 H_\uparrow$ and $\mu_0 H_\downarrow$ increase from 13 to 63 K and then decrease at higher temperatures, suggesting a maximum around 60 K. However, because of the limited temperature sampling and broad transformation ranges, the present data do not resolve a distinct maximum. In contrast, the more pronounced decrease at 156 K demonstrates a clear temperature dependence in this higher-temperature range. The history-induced reduction in the increasing-field transformation field, $D_{\uparrow,\mathrm{STC}} = \mu_0\left(H^{\mathrm{1st}}_{\uparrow,\mathrm{STC}} - H^{\mathrm{2nd}}_{\uparrow,\mathrm{STC}}\right)$, is approximately 13 T at 13, 63, and 97 K, which is smaller than $D_{\uparrow,\mathrm{ND}} =$ 18.6 T at 4.2 K. In contrast, the change in the decreasing-field transformation field, $D_{\downarrow,\mathrm{STC}} = \mu_0\left(H^{\mathrm{1st}}_{\downarrow,\mathrm{STC}} - H^{\mathrm{2nd}}_{\downarrow,\mathrm{STC}}\right)$, is much smaller in magnitude and is smaller than the corresponding transformation widths.

Figure 4 shows the hysteresis widths, $W = \mu_0(H_\uparrow - H_\downarrow)$, calculated from the transformation fields in Fig. 3. The hysteresis widths of the first application of STC pulse, $W^{\mathrm{1st}}_{\mathrm{STC}}$, remain approximately 22–25 T over the initial temperature range of 13–156 K, despite the changes in the individual transformation fields. The widths of the second field application, $W^{\mathrm{2nd}}_{\mathrm{STC}}$, are approximately 12 T at 13 and 63 K and 9.6 T at 97 K. These values correspond to a reduction of about 50–60% relative to the first field application. For ND at 4.2 K, the width decreases from $W^{\mathrm{1st}}_{\mathrm{ND}} =$ 22.2 T to $W^{\mathrm{2nd}}_{\mathrm{ND}} =$ 4.0 T, corresponding to a reduction of about 82%. Notably, the history-dependent suppression of transformation hysteresis persists under the ultrafast field sweeps of the STC system ($r \sim 10^7$ T/s), although the remaining hysteresis is larger than that measured using the ND system at 4.2 K ($r \sim 10^3$–$10^4$ T/s).

## Discussion

### Formation of a metastable martensitic state

The history dependence observed in the STC measurements is consistent with the interpretation proposed for the ND measurements [19]. Upon cooling in zero magnetic field, the sample transforms from A to M. Subsequent application of a magnetic field at low temperature induces the reverse martensitic transformation from M to A. When the field is removed, A transforms into M* rather than returning to the initial M. The first transformation cycle is therefore described as M → A → M*. Subsequently, during the second field application, the sample transforms between M* and A. Within this interpretation, the different martensitic starting states provide a natural explanation for the substantial change in $\mu_0 H_\uparrow$, whereas the common austenitic starting state is consistent with the much smaller change in

$\mu_0 H_\downarrow$.

During the negative-field portion of the first STC pulse, the martensite-to-austenite transformation occurs at a lower absolute field than during the initial field increase [Figs. 2(a) and 2(b)]. At an initial temperature of 13 K, this transformation occurs at −26.1 T, equal in magnitude to $\mu_0 H^{\mathrm{2nd}}_{\uparrow,\mathrm{STC}}$. Within the interpretation described above, this agreement indicates that M* had already formed during the preceding field removal on a microsecond timescale.

The history-induced reduction in $\mu_0 H_\uparrow$ is smaller in the STC measurements ($D_{\uparrow,\mathrm{STC}} \sim$ 13 T at 13, 63, and 97 K) than in the ND measurements ($D_{\uparrow,\mathrm{ND}} =$ 18.6 T at 4.2 K). Because $\mu_0 H^{\mathrm{1st}}_{\uparrow,\mathrm{STC}}$ at 13 K and $\mu_0 H^{\mathrm{1st}}_{\uparrow,\mathrm{ND}}$ at 4.2 K are nearly identical, the smaller $D_{\uparrow,\mathrm{STC}}$ mainly reflects the higher value of $\mu_0 H^{\mathrm{2nd}}_{\uparrow,\mathrm{STC}}$ compared with $\mu_0 H^{\mathrm{2nd}}_{\uparrow,\mathrm{ND}}$. The first and second increasing-field transformations correspond to M → A and M* → A, respectively. We next examine this contrast within the phenomenological framework described above, considering thermally activated and athermal contributions to the transformation field. The analysis takes into account the different sweep rates and initial temperatures, as well as possible differences in the metastable states formed under STC and ND fields.

**Thermally activated and athermal contributions**

The contrast between the first and second field applications can be discussed qualitatively using a phenomenological model based on Seeger's theory [21–23, 26–28]. In this model, the excess driving field required for transformation, $H_{\mathrm{ex}}$, is expressed as the sum of athermal and thermally activated contributions:

$$H_{\mathrm{ex}}(T,r) = H_\mu + H_{\mathrm{TA}}(T,r)$$

$$= H_\mu + H_{\mathrm{TA}}(0)\left[1-\left\{\frac{k_{\mathrm{B}}T}{Q_{0\mathrm{K}}}\ln\left(\frac{r_0}{r}\right)\right\}^{\frac{1}{q}}\right]^{\frac{1}{p}}. \quad (1)$$

Here, $H_\mu$ is the athermal contribution, assumed to be independent of temperature and field sweep rate for a fixed martensitic structure and microstructure. Its value may differ between the M → A and M* → A transformations. $H_{\mathrm{TA}}(0)$ is the zero-temperature limit of the thermally activated contribution, $H_{\mathrm{TA}}(T,r)$. $Q_{0\mathrm{K}}$ is the reference activation energy, $p$ and $q$ describe the barrier shape, which is estimated to be $p = 1/2$ and $q = 3/2$ for the case of MFIMT [22], $k_{\mathrm{B}}$ denotes the Boltzmann constant, and $r_0$ is a characteristic sweep-rate prefactor. The local sweep-rate magnitude is defined as $r = |d(\mu_0 H)/dt|$. Within this framework, the formation of M* may lower the transformation field through changes in $H_\mu$, $Q_{0\mathrm{K}}$, or both. Other

parameters including $H_{\mathrm{TA}}(0)$ may also depend on the martensite state.

For the increasing-field transformation, the measured transformation field is written as $H_{\uparrow} = H_0 + H_{\mu} + H_{\mathrm{TA}}(T, r)$, where $H_0$ is the equilibrium magnetic field between M and A for the first field application, and M* and A for the second field application. For each transformation, the difference between STC and ND measurements is $\Delta H_{\uparrow} = H_{\uparrow,\mathrm{STC}} - H_{\uparrow,\mathrm{ND}} = \Delta H_0 + \Delta H_{\mu} + \Delta H_{\mathrm{TA}}$. $H_0$ is determined by the thermodynamic states of the martensite and austenite. In the following discussion, $\Delta H_0$ is assumed to be negligible despite the difference in initial temperature and possible difference in the M* microstructure. However, differences in the twin microstructures of M* formed under STC and ND fields may lead to different $H_{\mu}$ even if their effect on $H_0$ is negligible. In this case, $\Delta H_{\mu}$ may contribute to the difference in the transformation field between the STC and ND measurements. The composition dependence of $H_{\mu}$ reported in Ni-Co-Mn-Al Heusler alloys indicates that $H_{\mu}$ is sensitive to material characteristics [29]. If $\Delta H_{\mu}$ is also negligible, we obtain $\Delta H_{\uparrow} \sim \Delta H_{\mathrm{TA}}$.

At a fixed temperature, slower field sweeps allow more time for thermal activation and reduce $H_{\mathrm{TA}}$. As the temperature approaches zero, this sweep-rate dependence weakens, and the excess driving field approaches the rate-independent limit: $\lim_{T\to 0} H_{\mathrm{ex}}(T) = H_{\mu} + H_{\mathrm{TA}}(0)$. The nearly identical values of $\mu_0 H_{\uparrow,\mathrm{STC}}^{\mathrm{1st}}$ at 13 K and $\mu_0 H_{\uparrow,\mathrm{ND}}^{\mathrm{1st}}$ at 4.2 K are compatible with the M → A transformation being close to this limit. In contrast, the higher value of $\mu_0 H_{\uparrow,\mathrm{STC}}^{\mathrm{2nd}}$ compared with $\mu_0 H_{\uparrow,\mathrm{ND}}^{\mathrm{2nd}}$ suggests that the M* → A transformation is more sensitive to the time available for thermal activation than the M → A transformation, provided that $H_0$ and $H_{\mu}$ are approximately the same between STC and ND measurements. It is noteworthy that, within the model, the higher temperature of the STC measurements reduces $H_{\mathrm{TA}}$ by promoting thermal activation, whereas the faster field sweep increases $H_{\mathrm{TA}}$ by allowing less time for thermal activation. Therefore, the higher $\mu_0 H_{\uparrow,\mathrm{STC}}^{\mathrm{2nd}}$ cannot be explained by the temperature mismatch alone and is consistent with the reduced thermal activation during the faster field sweep.

The finite time required for nucleation and subsequent interface motion may also influence the observed transformation field during STC pulses. The sweep-rate-dependent delay in the completion of the reverse martensitic transformation has been reported in Ni-Mn-In [30]. However, in the present study, the first-application transformation fields are nearly identical in STC and ND conditions. These results provide no clear evidence of a substantial delay in the M → A and A → M* transformations. Here, a delay specific to the M* → A transformation cannot be excluded at present.

Within the thermal activation framework, a smaller effective $Q_{0K}$ for the M* → A transformation provides a possible explanation for the difference between $\mu_0 H_{\uparrow,STC}^{2nd}$ and $\mu_0 H_{\uparrow,ND}^{2nd}$. If $r_0$ is comparable for the M → A and M* → A transformations, a smaller $Q_{0K}$ increases the dimensionless quantity $\frac{k_B T}{Q_{0K}} \ln\left(\frac{r_0}{r}\right)$ in Eq. (1) at a given temperature and sweep rate, allowing thermal activation to appreciably reduce $H_{TA}$ even at lower temperatures. In the regime close to the zero-temperature limit, this also increases the sweep-rate sensitivity if the other parameters are fixed. Thus, a lower activation barrier could explain both the lower transformation field and the stronger sweep-rate dependence. However, near the zero-temperature limit, a smaller $Q_{0K}$ also increases the temperature sensitivity of $H_{TA}$. Therefore, the weak temperature dependence of the hysteresis width observed at low temperatures in Fig. 4 and Ref. [19] requires further consideration.

Relaxation of M* toward M may explain the weak temperature dependence of the measured hysteresis width at low temperatures. Previous work proposed that M* can relax toward the 2M structure of M [19]. If this relaxation becomes faster with increasing temperature, the volume fraction of M present before the next field application may increase. This change may raise $\mu_0 H_\uparrow$, whereas thermal activation tends to lower it. Therefore, competition between thermal activation and changes in the relative volume fractions of M and M* could weaken the temperature dependence of $\mu_0 H_\uparrow$ and may contribute to the different temperature dependences of the hysteresis width observed in the STC and ND measurements. The hysteresis width measured during STC pulses shows a weak temperature dependence between 13 and 97 K. In contrast, the previously reported hysteresis width measured during ND pulses is approximately constant below 60 K, increases as the temperature increases from 60 to 180 K, and decreases above 180 K [19].

Alternatively, differences in the twin microstructure of M* formed under STC and ND fields may lead to different $H_\mu$ values. Such differences in $H_\mu$ may contribute to the difference between $\mu_0 H_{\uparrow,STC}^{2nd}$ and $\mu_0 H_{\uparrow,ND}^{2nd}$. In Eq. (1), $H_\mu$ is independent of temperature and field sweep rate for a fixed martensite state. However, the twin microstructure of M* and the relative volume fraction of M and M* may depend on temperature and field sweep rate. Therefore, changes in these structural characteristics could produce a temperature- and sweep-rate-dependent effective athermal contribution.

The internal structure of M* may affect both the effective activation energy $Q_{0K}$ and the athermal contribution $H_\mu$. Theoretical studies of $Ni_2MnGa$ predict that transformation pathways involving modulated martensitic structures, such as 4O, 6M, 10M, or 14M, can reduce the transformation barrier [31–33]. In addition, reduced thermal transformation hysteresis associated with the formation of 14M martensite has been observed experimentally in $Ni_{51.9}Mn_{27}Ga_{21.1}$ [20]. Differences in stacking sequences or twin microstructure may also change how the transformation strain is accommodated at the austenite–martensite interface, potentially modifying $Q_{0K}$ and $H_{\mu}$. However, the present magnetization measurements cannot determine the relative contributions of changes in $Q_{0K}$ and $H_{\mu}$ or directly identify the structure of $M^*$. Therefore, time-resolved X-ray diffraction during high-field cycling is crucial to identify the structure of $M^*$ formed during MFIMTs.

**Conclusions**

Magnetization measurements of $Ni_{41}Co_9Mn_{31.5}Ga_{18.5}$ using microsecond STC pulses show that magnetic history reduces the transformation hysteresis width by approximately 50–60% at initial temperatures of 13–97 K. This reduction is dominated by a decrease in the increasing-field transformation field. The reduced absolute transformation field observed during the negative-field portion of the first STC pulse further supports the formation of a metastable martensitic state on a microsecond timescale during field removal.

Comparison with ND measurements suggests that the increasing-field transformation field is more sensitive to the magnetic-field sweep rate for $M^* \rightarrow A$ than for $M \rightarrow A$. Within the phenomenological framework, a lower effective activation barrier for $M^* \rightarrow A$ could enhance the thermal activation and contribute to the stronger sweep-rate sensitivity of its transformation field. Alternatively, differences in the twin microstructure of $M^*$ formed under STC and ND fields may produce different athermal contributions $H_{\mu}$. Relaxation of $M^*$ toward M at higher temperatures may increase the transformation field, partly offsetting its reduction by thermal activation. This competition could weaken the temperature dependence of the increasing-field transformation field. Time-resolved X-ray diffraction measurements during and after magnetic field cycling are needed to identify the crystal structure and twin microstructure of $M^*$ and its contribution to the activation barrier and $H_{\mu}$.

**Acknowledgment**

This work was supported in part by the MEXT Leading Initiative for Excellent Young Researchers (Grant No. JPMXS0320210045), JSPS KAKENHI Grant Numbers JP23K26366, JP26K01195, JP24K21225, and JP24K17003, and the Foundation of Kinoshita Memorial

Enterprise.

## References

1. Pecharsky VK, Gschneidner KA (1997) Giant magnetocaloric effect in $Gd_5(Si_2Ge_2)$. Phys Rev Lett 78:4494–4497. https://doi.org/10.1103/PhysRevLett.78.4494
2. Morellon L, Algarabel PA, Ibarra MR, Blasco J, Garcí A-Landa B, Arnold Z, Albertini F (1998) Magnetic-field-induced structural phase transition in $Gd_5(Si_{1.8}Ge_{2.2})$. Phys Rev B 58:R14721. https://doi.org/10.1103/PhysRevB.58.R14721
3. Umetsu RY, Xu X, Kainuma R (2016) NiMn-based metamagnetic shape memory alloys. Scr Mater 116:1–6. https://doi.org/10.1016/j.scriptamat.2016.01.006
4. Kihara T, Roy T, Xu X, Miyake A, Tsujikawa M, Mitamura H, Tokunaga M, Adachi Y, Eto T, Kanomata T (2021) Observation of inverse magnetocaloric effect in magnetic-field-induced austenite phase of Heusler alloys $Ni_{50-x}Co_xMn_{31.5}Ga_{18.5}$ (x = 9 and 9.7). Phys Rev Mater 5:034416. https://doi.org/10.1103/PhysRevMaterials.5.034416
5. Matsuda YH, Muraoka Y, Nakamura D, Ikeda A, Ishii Y, Zhou X-G, Sawabe H, Takeyama S (2022) Magnetic-Field-Induced Insulator Metal Transition of W-doped $VO_2$ Observed by Electromagnetic Flux Compression at ISSP. J Phys Soc Jpn 91:101008. https://doi.org/10.7566/jpsj.91.101008
6. Matsuda YH, Nakamura D, Ikeda A, Takeyama S, Suga Y, Nakahara H, Muraoka Y (2020) Magnetic-field-induced insulator–metal transition in W-doped $VO_2$ at 500 T. Nature Communications 11:3591. https://doi.org/10.1038/s41467-020-17416-w
7. Ikeda A, Matsuda YH, Sato K (2020) Two Spin-State Crystallizations in $LaCoO_3$. Phys Rev Lett 125:177202. https://doi.org/10.1103/PhysRevLett.125.177202
8. Ikeda A, Kubota Y, Ishii Y, Zhou X, Peng S, Hayashi H, Matsuda YH, Noda K, Tanaka T, Shimbori K, Seki K, Kobayashi H, Bhoi D, Gen M, Gautam K, Akaki M, Kawachi S, Kasamatsu S, Nomura T, et al (2025) X-Ray Free-Electron Laser Observation of Giant and Anisotropic Magnetostriction in $\beta$-$O_2$ at 110 Tesla. Phys Rev Lett 135:186702. https://doi.org/10.1103/r7br-qnrn
9. Nomura T, Matsuda YH, Takeyama S, Matsuo A, Kindo K, Her JL, Kobayashi TC (2014) Novel phase of solid oxygen induced by ultrahigh magnetic fields. Phys Rev Lett 112:247201. https://doi.org/10.1103/PhysRevLett.112.247201
10. Kainuma R, Imano Y, Ito W, Sutou Y, Morito H, Okamoto S, Kitakami O, Oikawa K, Fujita A, Kanomata T, Ishida K (2006) Magnetic-field-induced shape recovery by reverse phase transformation. Nature 439:957–960. https://doi.org/10.1038/nature04493
11. Kihara T, Xu X, Ito W, Kainuma R, Tokunaga M (2014) Direct measurements of inverse magnetocaloric effects in metamagnetic shape-memory alloy NiCoMnIn. Phys Rev B 90:214409. https://doi.org/10.1103/PhysRevB.90.214409
12. Liu J, Gottschall T, Skokov KP, Moore JD, Gutfleisch O (2012) Giant magnetocaloric

effect driven by structural transitions. Nat Mater 11:620–626. https://doi.org/10.1038/nmat3334

13. Krenke T, Duman E, Acet M, Wassermann EF, Moya X, Mañosa L, Planes A (2005) Inverse magnetocaloric effect in ferromagnetic Ni-Mn-Sn alloys. Nat Mater 4:450–454. https://doi.org/10.1038/nmat1395

14. Guillou F, Pathak AK, Paudyal D, Mudryk Y, Wilhelm F, Rogalev A, Pecharsky VK (2018) Non-hysteretic first-order phase transition with large latent heat and giant low-field magnetocaloric effect. Nat Commun 9:2925. https://doi.org/10.1038/s41467-018-05268-4

15. Gutfleisch O, Gottschall T, Fries M, Benke D, Radulov I, Skokov KP, Wende H, Gruner M, Acet M, Entel P, Farle M (2016) Mastering hysteresis in magnetocaloric materials. Philos Trans R Soc A 374:20150308. https://doi.org/10.1098/rsta.2015.0308

16. Scheibel F, Gottschall T, Taubel A, Fries M, Skokov KP, Terwey A, Keune W, Ollefs K, Wende H, Farle M, Acet M, Gutfleisch O, Gruner ME (2018) Hysteresis Design of Magnetocaloric Materials—From Basic Mechanisms to Applications. Energy Technol 6:1397–1428. https://doi.org/10.1002/ente.201800264

17. Cui J, Chu YS, Famodu OO, Furuya Y, Hattrick-Simpers J, James RD, Ludwig A, Thienhaus S, Wuttig M, Zhang Z, Takeuchi I (2006) Combinatorial search of thermoelastic shape-memory alloys with extremely small hysteresis width. Nat Mater 5:286–290. https://doi.org/10.1038/nmat1593

18. Sepehri-Amin H, Taubel A, Ohkubo T, Skokov KP, Gutfleisch O, Hono K (2018) Microstructural origin of hysteresis in Ni-Mn-In based magnetocaloric compounds. Acta Mater 147:342–349. https://doi.org/10.1016/j.actamat.2018.01.044

19. Kihara T, Xu X, Miyake A, Kinoshita Y, Tokunaga M, Adachi Y, Kanomata T (2020) High-field magnetization and magnetic phase diagram of metamagnetic shape memory alloys $Ni_{50-x}Co_xMn_{31.5}Ga_{18.5}$ ( x = 9 and 9.7). Scr Mater 181:25–29. https://doi.org/10.1016/j.scriptamat.2020.02.009

20. Konoplyuk SM, Kolomiets A, Daniš S, Míšek M, Kokorin V V., Vondračková B, Kozlova LE (2018) Origin of temperature hysteresis of the magnetic transition in martensitic phase of $Ni_{51.9}Mn_{27}Ga_{21.1}$ alloy. Scr Mater 154:60–63. https://doi.org/10.1016/j.scriptamat.2018.05.031

21. Seeger A (1954) The temperature dependence of the critical shear stress and of work-hardening of metal crystals. Edinburgh Dublin Philos Mag J Sci 45:771–773. https://doi.org/10.1080/14786440708520489

22. Umetsu RY, Endo K, Kanomata T, Xu X, Kainuma R, Kondo A, Kindo K, Ito W (2013) Magnetoresistance and transformation hysteresis in the Ni $_{50}Mn_{34.4}In_{15.6}$ metamagnetic shape memory alloy. Mater Trans 54:291–296. https://doi.org/10.2320/matertrans.M2012365

23. Xu X, Kihara T, Tokunaga M, Matsuo A, Ito W, Umetsu RY, Kindo K, Kainuma R (2013) Magnetic field hysteresis under various sweeping rates for Ni-Co-Mn-In metamagnetic shape memory alloys. Appl Phys Lett 103:122406. https://doi.org/10.1063/1.4821184
24. Nayak AK, Suresh KG, Nigam AK (2011) Metastability of magneto-structural transition revealed by sweep rate dependence of magnetization in $Ni_{45}Co_{5}Mn_{38}Sb_{12}$ Heusler alloy. J Appl Phys 109:07A906. https://doi.org/10.1063/1.3536601
25. Umetsu RY, Xu X, Ito W, Kihara T, Takahashi K, Tokunaga M, Kainuma R (2014) Magnetic Field-Induced reverse martensitic transformation and thermal transformation arrest phenomenon of Ni41Co9Mn39Sb11 alloy. Metals (Basel) 4:609–622. https://doi.org/10.3390/met4040609
26. Ghosh G, Olson GB (1994) Kinetics of F.C.C. → B.C.C. Heterogeneous Martensitic Nucleation-II. Thermal Activation. Acta Metall Mater 42:3371–3379. https://doi.org/10.1016/0956-7151(94)90469-3
27. Ghosh G, Olson GB (1994) Kinetics of F.C.C. → B.C.C. Heterogeneous Martensitic Nucleation-I.The Critical Driving Force for Athermal Nucleation. Acta Metall Mater 42:3361–3370. https://doi.org/10.1016/0956-7151(94)90468-5
28. Kocks UF, Argon AS, Ashby MF (1975) Thermodynamics and Kinetics of Slip. Prog Mater Sci 19:1
29. Xu X, Ito W, Tokunaga M, Kihara T, Oka K, Umetsu RY, Kanomata T, Kainuma R (2013) The thermal transformation arrest phenomenon in NiCoMnAl heusler alloys. Metals (Basel) 3:298–311. https://doi.org/10.3390/met3030298
30. Pfeuffer L, Gottschall T, Faske T, Taubel A, Scheibel F, Karpenkov AY, Ener S, Skokov KP, Gutfleisch O (2020) Influence of the martensitic transformation kinetics on the magnetocaloric effect in Ni-Mn-In. Phys Rev Mater 4:111401(R). https://doi.org/10.1103/PhysRevMaterials.4.111401
31. Zelený M, Straka L, Sozinov A, Heczko O (2018) Transformation Paths from Cubic to Low-Symmetry Structures in Heusler $Ni_2MnGa$ Compound. Sci Rep 8:7275. https://doi.org/10.1038/s41598-018-25598-z
32. Zelený M, Sedlák P, Heczko O, Seiner H, Veřtát P, Obata M, Kotani T, Oda T, Straka L (2021) Effect of electron localization in theoretical design of Ni-Mn-Ga based magnetic shape memory alloys. Mater Des 209:109917. https://doi.org/10.1016/j.matdes.2021.109917
33. Gruner ME, Niemann R, Entel P, Pentcheva R, Rößler UK, Nielsch K, Fähler S (2018) Modulations in martensitic Heusler alloys originate from nanotwin ordering. Sci Rep 8:8489. https://doi.org/10.1038/s41598-018-26652-6

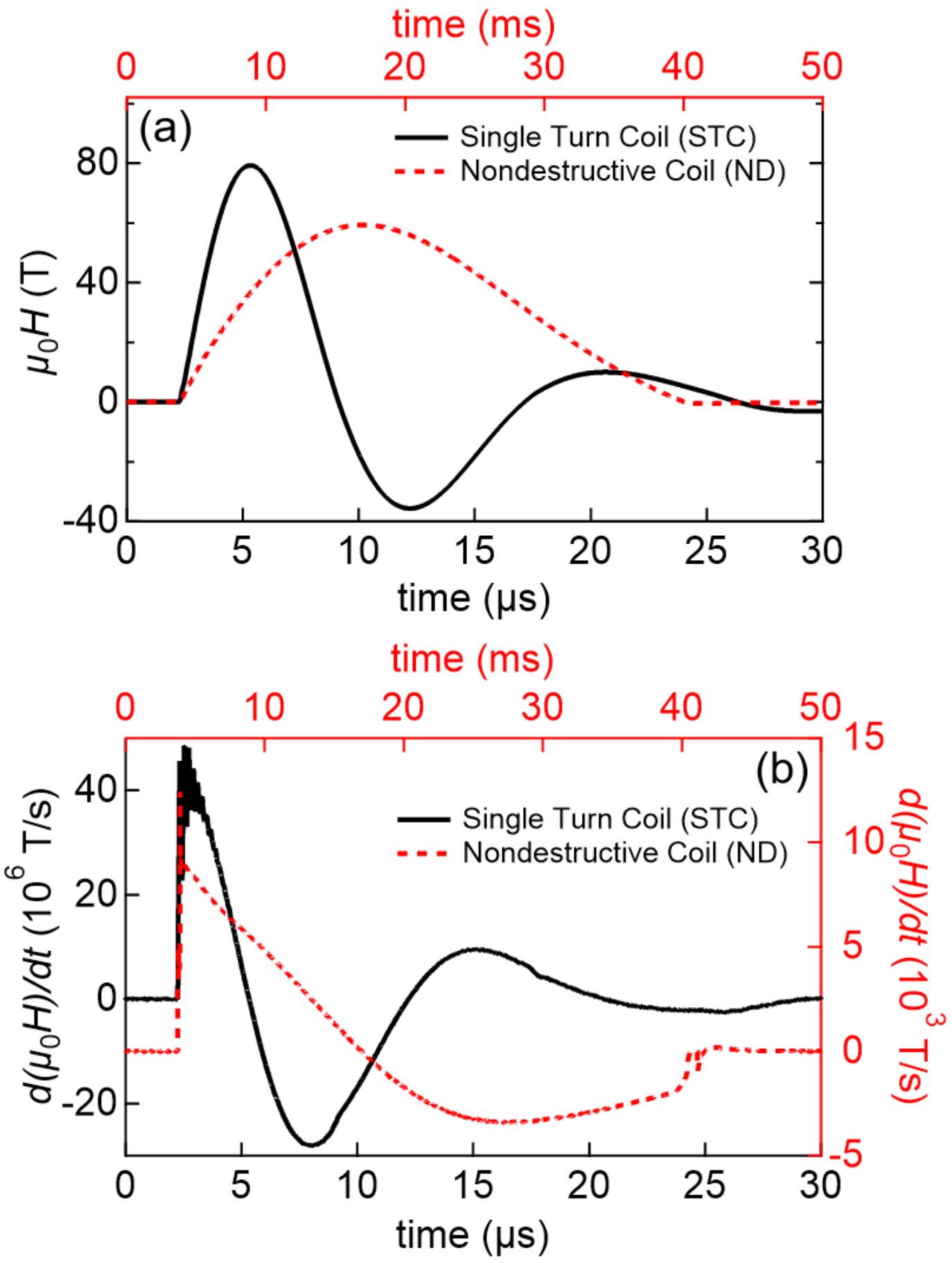

Fig. 1 (a) Representative magnetic field profiles plotted as functions of time for the single-turn-coil (STC, solid black curve, bottom time axis) and nondestructive (ND [19], dashed red curve, top time axis) pulsed field systems. (b) Corresponding magnetic field sweep rates, $d(\mu_0 H)/dt$ .

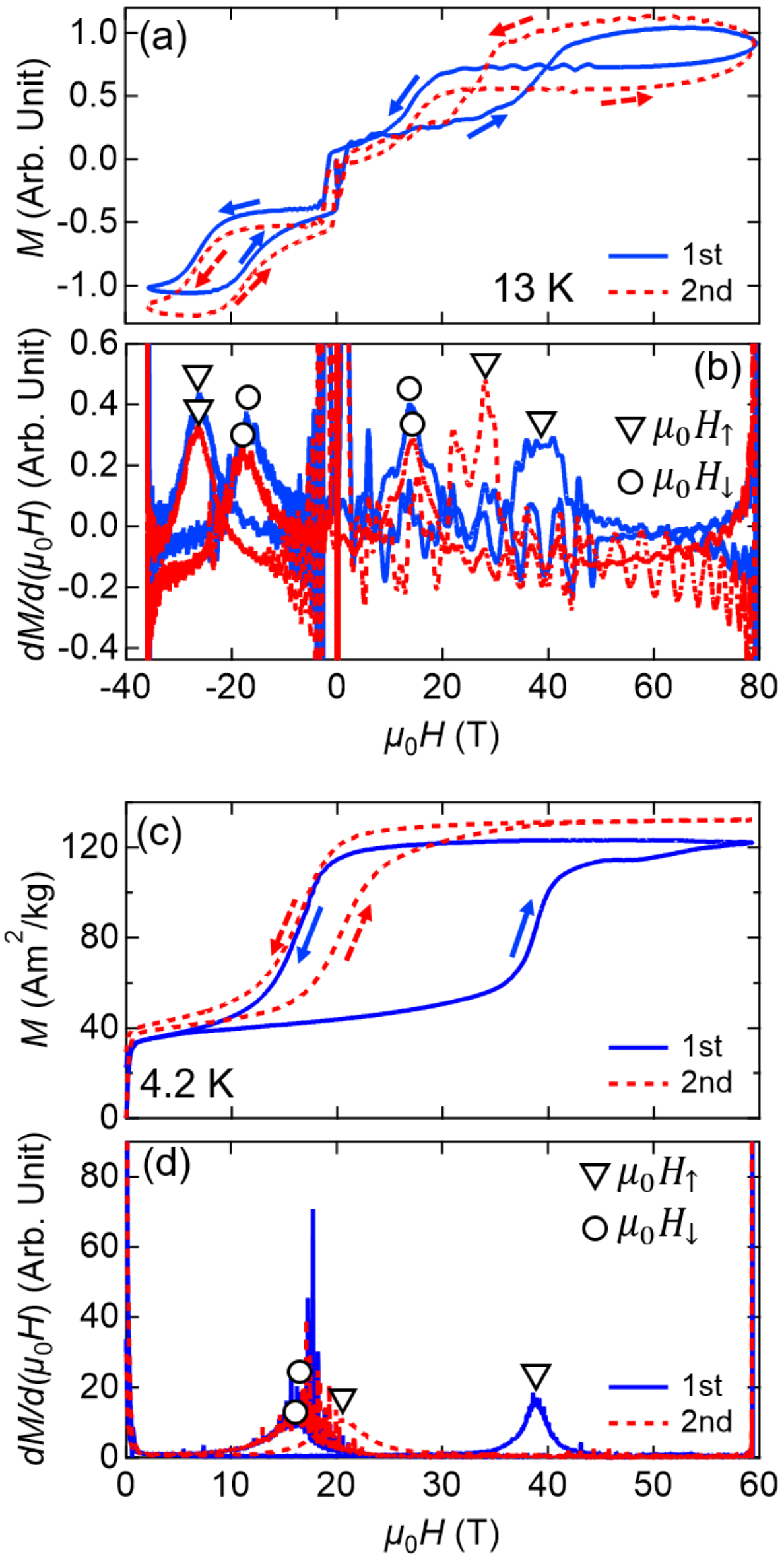

Fig. 2 Magnetization curves and their field derivatives for $Ni_{41}Co_9Mn_{31.5}Ga_{18.5}$ measured using (a, b) the STC system at 13 K and (c, d) the ND system at 4.2 K. Solid blue and dashed red curves represent the first and second field applications, respectively. Open inverted triangles and circles mark the transformation fields of the martensite-to-austenite and austenite-to-martensite, respectively, determined from peaks in $dM/d(\mu_0 H)$.

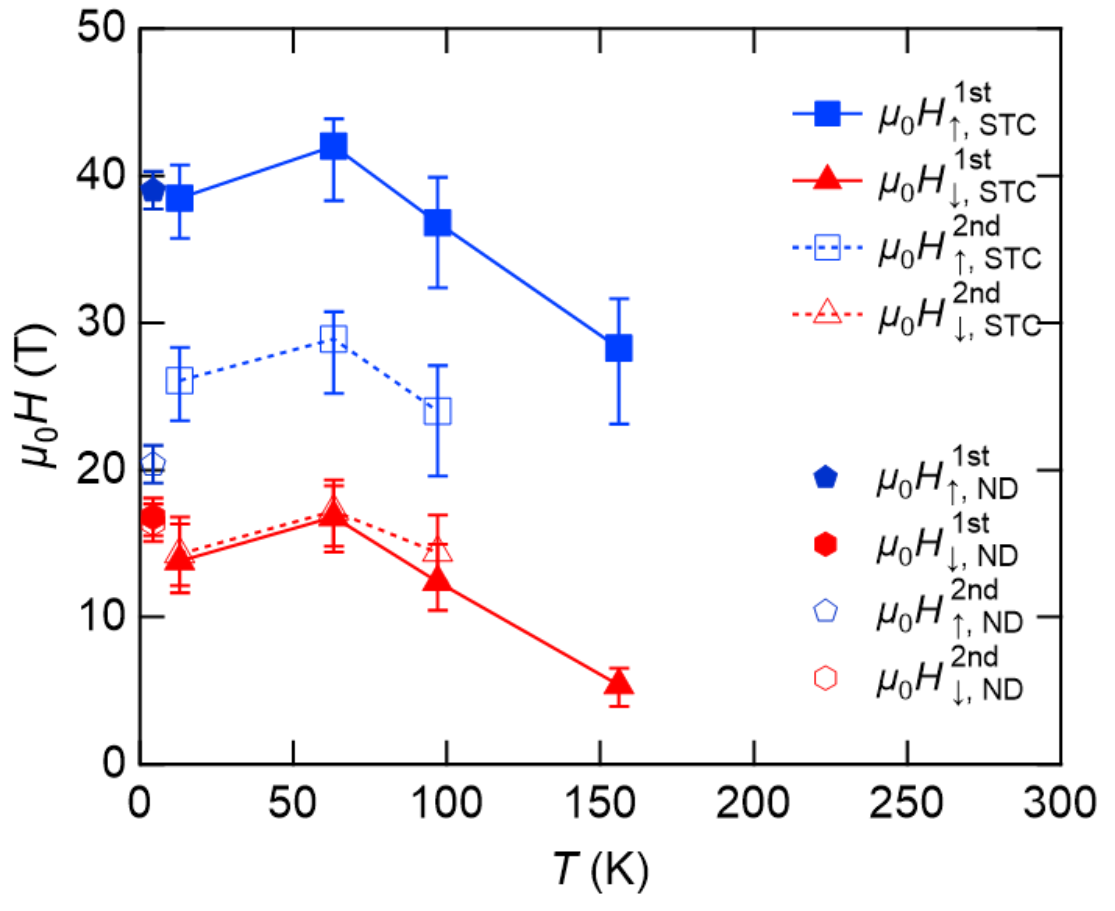

Fig. 3 Temperature dependence of the transformation fields of $Ni_{41}Co_9Mn_{31.5}Ga_{18.5}$ measured using the STC and ND systems. Blue and red symbols denote the transformation fields on the field-increasing, $H_\uparrow$, and field-decreasing, $H_\downarrow$, processes, respectively. Filled and open symbols represent the first and second field applications. Squares and triangles show the STC results, and pentagons and hexagons show the ND results at 4.2 K. The bars indicate the widths of the corresponding peaks in $dM/d(\mu_0 H)$.

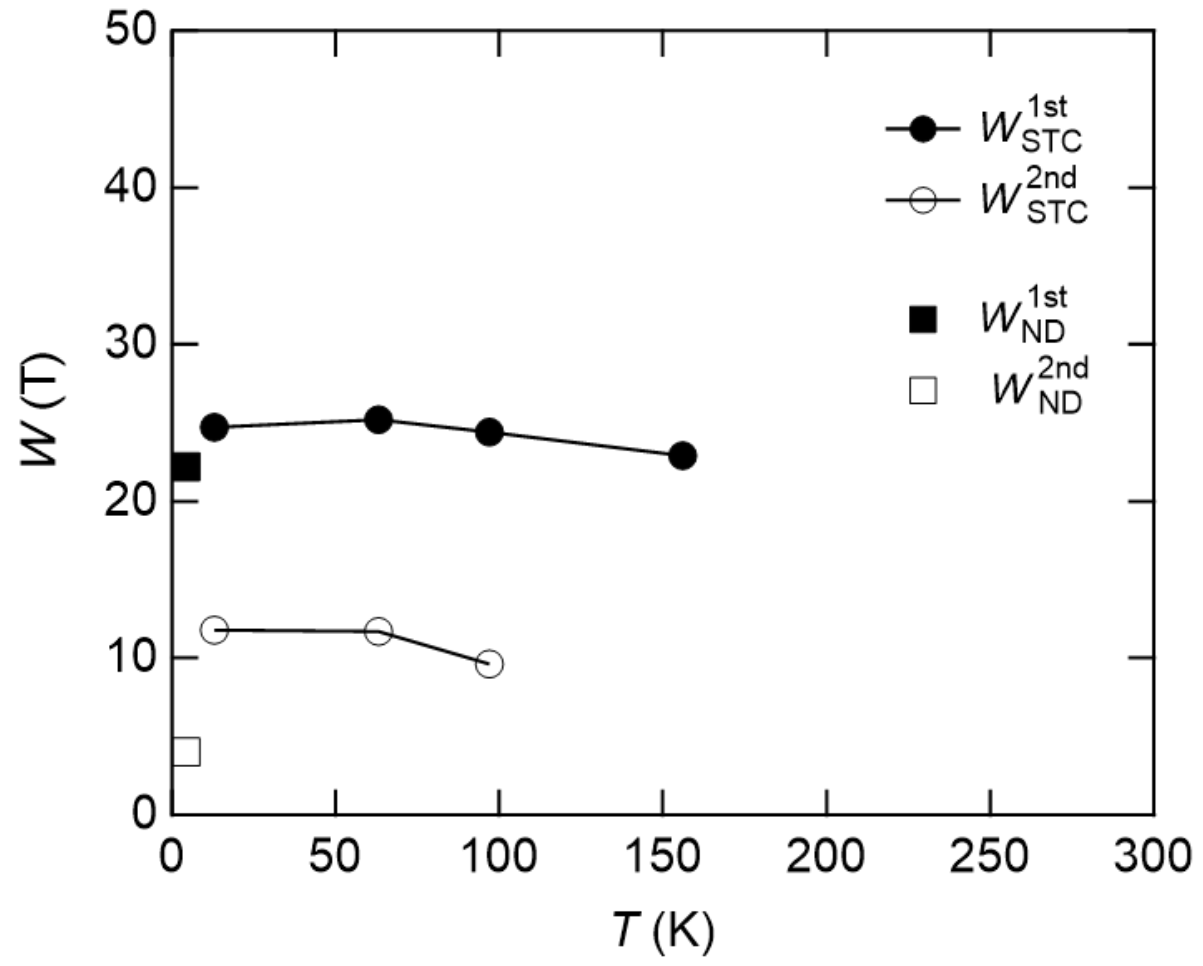

Fig. 4 Temperature dependence of the transformation hysteresis width, $W = \mu_0(H_\uparrow - H_\downarrow)$, calculated from the transformation fields in Fig. 3. Circles and squares represent the STC and ND results, respectively. Filled and open symbols denote the first and second field applications.